\documentclass[a4paper,12pt]{article}
\usepackage{tikz} 
\usepackage{graphicx}
 
\usepackage{mathtools} 
\usepackage{braket}
\usepackage{xcolor}
\usepackage[utf8]{inputenc}
\usepackage[english]{babel}
\usepackage{amsmath,amsfonts,amssymb}
\usepackage[scaled]{helvet}
\usepackage{url}
\usepackage[colorlinks=true, allcolors=blue]{hyperref}
\usepackage{authblk}
\usepackage{xifthen}
\usepackage{textcomp} 
\usepackage{graphicx,xcolor}
\usepackage{wrapfig} 
\usepackage{colortbl}
\usepackage{booktabs}
\usepackage{algorithm}
\usepackage[noend]{algpseudocode}
\usepackage{changepage}
\usepackage[numbers,sort&compress]{natbib}
\usepackage{fancyhdr}  
\usepackage{lastpage}  
\usepackage[explicit]{titlesec}
\usepackage{enumitem} 
\usepackage[resetlabels]{multibib}
\usepackage{letltxmacro}
\usepackage{xparse}
\usepackage{silence}
\usepackage{xpatch}
\usepackage{csquotes}
\usepackage[font=small,labelfont=bf, margin=1cm]{caption}

\usepackage[T1]{fontenc}
\usepackage{empheq}
\usepackage{siunitx}
\usepackage{soul}
\usepackage{pifont}

\usetikzlibrary{shapes.arrows, fadings}
\tikzfading[name=fade below, top color=transparent!0, bottom color=transparent!100]

\def\curl{\ensuremath{\mathbf{curl}}\,}
\def\div{\mathrm{div}\,}
\def\grad{\ensuremath{\mathbf{grad}}\,}

\def\bE{\ensuremath{\mathbf{E}}} 
\def\bS{\ensuremath{\mathbf{S}}}
\def\be{\ensuremath{\mathbf{e}}}
\def\bh{\ensuremath{\mathbf{h}}}
\def\bW{\ensuremath{\overline{\mathbf{W}}}}

\def\bn{\ensuremath{\mathbf{n}}}
\def\bx{\ensuremath{\mathbf{r}}}
\def\bx{\ensuremath{\mathbf{x}}}

\def\bzh{\ensuremath{\hat{\mathbf{z}}}}

\def\epsrdd{\ensuremath\varepsilon_{r,2D}}
\def\epsrddd{\ensuremath\varepsilon_{r,3D}}
\def\tensmurdd{\ensuremath\boldsymbol{\mu}_{r,2D}}
\def\tensmurddd{\ensuremath\boldsymbol{\mu}_{r,3D}}
\def\tensepsrdd{\ensuremath\boldsymbol{\varepsilon}_{r,2D}}
\def\tensepsrddd{\ensuremath\boldsymbol{\varepsilon}_{r,3D}}
\def\tensmurp{\ensuremath\boldsymbol{\mu}_{r,\#}}
\def\tensepsrp{\ensuremath\boldsymbol{\varepsilon}_{r,\#}}
\newcommand{\epsrin}[1]{\ensuremath\varepsilon_{r,#1}}
\newcommand{\mic}[1]{\SI{#1}{\micro\metre}}

\def\om{\ensuremath{\omega}}

\def\betak{\ensuremath{\beta_k}}
\def\d{\ensuremath{\,\mathrm{d}}}

\def\x{\mathbf{x}}

\def\dint{\displaystyle\int}

\def\tensid{\underline{\underline{I}}}

\def\bE{\ensuremath{\mathbf{E}}}
\def\bH{\ensuremath{\mathbf{H}}}
\def\bEp{\mathrm{\mathbf{E}}_{\#}}
\def\bEpk{\mathrm{\mathbf{E}}_{\#,k}}
\def\bEinc{\mathrm{\mathbf{E}}^{\mathrm{inc}}}
\def\bHinc{\mathrm{\mathbf{H}}^{\mathrm{inc}}}

\def\bek{\ensuremath{\mathbf{e}_{k}}}

\def\bEd{\mathrm{\mathbf{E}}^{\mathrm{d}}}
\def\bHd{\mathrm{\mathbf{H}}^{\mathrm{d}}}
\def\bEtot{\mathrm{\mathbf{E}}^{\mathrm{tot}}}
\def\bStot{\mathrm{\mathbf{S}}^{\mathrm{tot}}}
\def\bSi{\mathrm{\mathbf{S}}^{\mathrm{inc}}}
\def\bSd{\mathrm{\mathbf{S}}^{\mathrm{d}}} 
\def\bHtot{\mathrm{\mathbf{H}}^{\mathrm{tot}}}

\newcommand{\real}[1]{\ensuremath{\mathrm{Re}\{#1\}}}
\newcommand{\bEdd}[1]{\ensuremath{\mathbf{E}_{#1,2D}}}

\definecolor{fresneldarkblue}{RGB}{44,46,131} 
\definecolor{colorguide}{RGB}{154, 217, 255} 
\definecolor{fresnellightblue}{RGB}{0,165,226} 
\definecolor{fresnelgrey}{RGB}{111,111,110} 
\definecolor{colorrod}{RGB}{213, 188, 127} 
\definecolor{colorlow}{RGB}{73, 229, 140} 
\definecolor{colorsubs}{RGB}{153, 111, 154} 
\definecolor{colorrewier1}{RGB}{66, 134, 244}
\definecolor{colorrewier2}{RGB}{244, 110, 66}
\definecolor{colorrewier3}{RGB}{55, 166, 55}
\definecolor{colorme}{RGB}{186, 52, 235}
\def\crevi{\color{black}}
\def\crevii{\color{black}}

\def\crevme{\color{black}}

\title{Discontinuities in photonic waveguides: Rigorous Maxwell-based 3D modeling with the finite element method} 

\author{Guillaume Demésy\footnote{Corresponding author : \texttt{guillaume.demesy@fresnel.fr}}~~and Gilles Renversez\\
\footnotesize{Aix Marseille Univ, CNRS, Centrale Marseille, Institut Fresnel, Marseille, France.}}

\begin{document}

\maketitle
\begin{abstract}
  In this paper, a general methodology to study rigorously  discontinuities in open waveguides is presented. It relies on a full vector description given by Maxwell's equations in the framework of the finite element method. The discontinuities are not necessarily small perturbations of the initial waveguide and can be very general, such as plasmonic inclusions of arbitrary shapes. The leaky modes of the invariant structure {\crevme are first computed and then injected} as incident fields in the full structure with obstacles using a scattered field approach. The resulting scattered field is finally projected on the modes of the invariant structure making use of their bi-orthogonality. The energy balance is discussed. Finally, the modes of open waveguides periodically structured along the propagation direction are computed. The relevant complex propagation constants are compared to the transmission obtained for a finite number of identical cells. The relevance and complementarity of the two approaches are highlighted on a numerical example encountered in infrared sensing. Open source models allowing to retrieve most of the results of this paper are provided.
\end{abstract}

\section{Introduction}
\label{sec:intro}
The study of discontinuities is an old research topic in waveguide studies due to its importance for practical applications in many areas of physics. One must cite the seminal contribution of Schwinger for the development of variational methods in the forties~\cite{schinger68discontinuities-waveguides} and the results obtained by Lewin~\cite{lewin75theory-waveguides}. 

These methods, often complex and specific, do not generally consider the exact solutions of Maxwell's equations and rely on specific configurations, hypotheses, initial guesses for the solution forms. During the last two decades, the versatile Finite-Difference Time-Domain (FDTD) method allowed the study of waveguide discontinuities, including 3D ones, taking into account the full set of Maxwell's equations~\cite{taflove05,taflove13}.
Nevertheless, the computational resources both in terms of memory and time requirements are huge when realistic 3D photonic devices are considered with a uniform square grid, especially nanophotonic ones with high quality factors. As for harmonic methods, Fourier modal methods \cite{lalanne2000fourier,lecamp2007theoretical} also allow to tackle discontinuities in waveguides but they are restricted to geometries with straight walls. In acoustics and optics, coupled modal-finite element techniques have been successfully used in varying cross-section waveguides~\cite{pagneux1996study,pelat2011coupled}. 

{\crevii Two types of non-Hermitian eigenvalue problems arise in open nanophotonic structures. When considering resonators in the general case, no particular ansatz can be guessed for the electromagnetic field and a natural eigenvalue is the (complex) frequency. Diffraction gratings are a special yet frequent case where the Bloch theorem applies ; the Bloch variable arises as a wavenumber and a corner stone of the dispersion relation of gratings consists in looking for \emph{complex frequencies for a given real wavenumber}. Recent benchmarks of the numerical methods cited in the previous paragraph can be found in Refs.~\cite{rosenkrantz2018benchmarking,lalanne2019quasinormal} for this type of non-Hermitian eigenvalue problems. But when considering guiding structures as it is the case in this article, a more natural eigenvalue is \emph{the propagation constant at a given real frequency}. When coupling a laser -- indeed operating at a real frequency~-- into a waveguide, the relavant quantities are the light velocity and attenuation which are directly related to the real and imaginary parts of the eigenvalues (\emph{i.e.} complex propagation constants) of the modes of the leaky waveguide.}

This article addresses the numerical characterization of open waveguides with discontinuities. We demonstrate that it can be carried out efficiently with adapted formulations of the finite element method (FEM) which has already proven its efficiency and versatility in many field of computational electrodynamics~\cite{jin02FEM-electromag}.

We can state four main advantages of the FEM-based method: i) curved geometries are naturally treated using high order mesh elements and corresponding shape functions, ii) conforming non-uniform meshing is now a standard for mesh generators which is particularly relevant when rapid and strong permittivity changes must be tackled, iii) the domain decomposition method, now available in several FEM solvers, allows the treatment of large scale 3D problem, iv) and the possibility to reuse the inverse matrix for several incident modes propagating in the invariant structure -- a subtlety detailed later which is a key advantage for the optimization multi-mode guides. This is especially worthy when the simulations are performed within a topology optimization frame~\cite{bendoe-sigmund04topo-opti}.

The practical context motivating this theoretical and numerical study is the design of efficient plasmonic waveguides for infrared sensing~\cite{gutierrez2016optical} since the mid-IR spectral domain is known  to be the molecular fingerprint region, due to the fact that most molecule including pollutants have intense fundamental vibrational bands in this spectral range. The device configuration is fully integrated and based on a ridge waveguide upon which metallic scattering nano-objects will ensure the coupling between the guided modes and superstrate of the device. Chalcogenide glasses are chosen for the main layers due to their high transparencies for infrared wavelengths \cite{Desevedavy:09,eggleton2011chalcogenide}. Ultimately, the metallic scatterers are planned to be functionalized in order to react to the targeted chemical species. The sensing property relies on the subsequent modification of the guidance of the full structure. 

{\crevme With this application in mind,} we present a general framework to study rigorously  discontinuous waveguides using a full vector description given by Maxwell's equations in the framework of the finite element method. The discontinuity can be very general and is not necessarily a small perturbation of the initial waveguide. The full structure under investigation is made of 3 segments: The input one is a uniform waveguide invariant along its main propagation axis, the intermediate one {\crevme (called ``modified segment'' in Fig.~\ref{fig:scheme})} contains the opto-geometrical modifications of the waveguide, the output one is again an invariant waveguide. In order to model the response of the resulting 3D guiding structure, we adopt a scattered field formulation consisting of three sequential steps, the output of first step being the input of the second one, the output of the first and  second being the input of the third one.

First, we determine the leaky modes of the unperturbed 2D waveguide for a fixed frequency corresponding to the freespace wavelength of interest. {\crevme The example chosen to illustrate our method consists in} a ridge waveguide made of chalcogenide layers on a silicon substrate, assumed to be invariant along its propagation axis. We use our usual vector FEM method with the Galerkin approach to solve the relevant eigenvalue problem~\cite{nedelec91,jin02FEM-electromag,renversez2012foundations}. This first step provides both the propagation constants (eigenvalues) and the associated modes profiles (eigenvectors).

Second, these guided modes are used as incident fields for the full 3D problem in the modified segment. The electromagnetic problem to solve for this second step is then a mere scattering problem~\cite{demesy2010all}. It is possible to define a proper energy balance (transmission and reflection, absorption taking place into the obstacles, extra radiation losses) allowing to fully evaluate the impact of the modified segment on the energy propagation. 

Third, outside of the modified region, the total field is {\crevme expanded} on a fixed number of leaky modes of the output segment of the full structure. {\crevi A special attention is paid to the coupling efficiency into the mode initially injected after crossing the modified segment.} Our method allows to compute all the required energy-related quantities to investigate quantitatively the behavior of the full structure, notably the impact of the modified segment, and to take into account the way it is excited by the selected input propagating mode.

Note that our approach differs from the one exposed in Ref.~\cite{jin02FEM-electromag} where total field formulations making use of port boundary conditions are applied to closed discontinuous waveguides, whereas it is proposed here to use a general scattered field formulation to deal with open discontinuous waveguides.

Finally, we also compute the modes of waveguides infinitely periodically structured along the propagation axis and compare the relevant complex propagation constants to the transmission obtained with a finite number of identical cells. After deriving the formulation, the relevance and complementarity of the two approaches are highlighted on a numerical example.

\section{Direct problem}\label{sec:3Ddirect} 
\begin{figure}[t] 
  \centering
  \includegraphics[width=.7\linewidth]{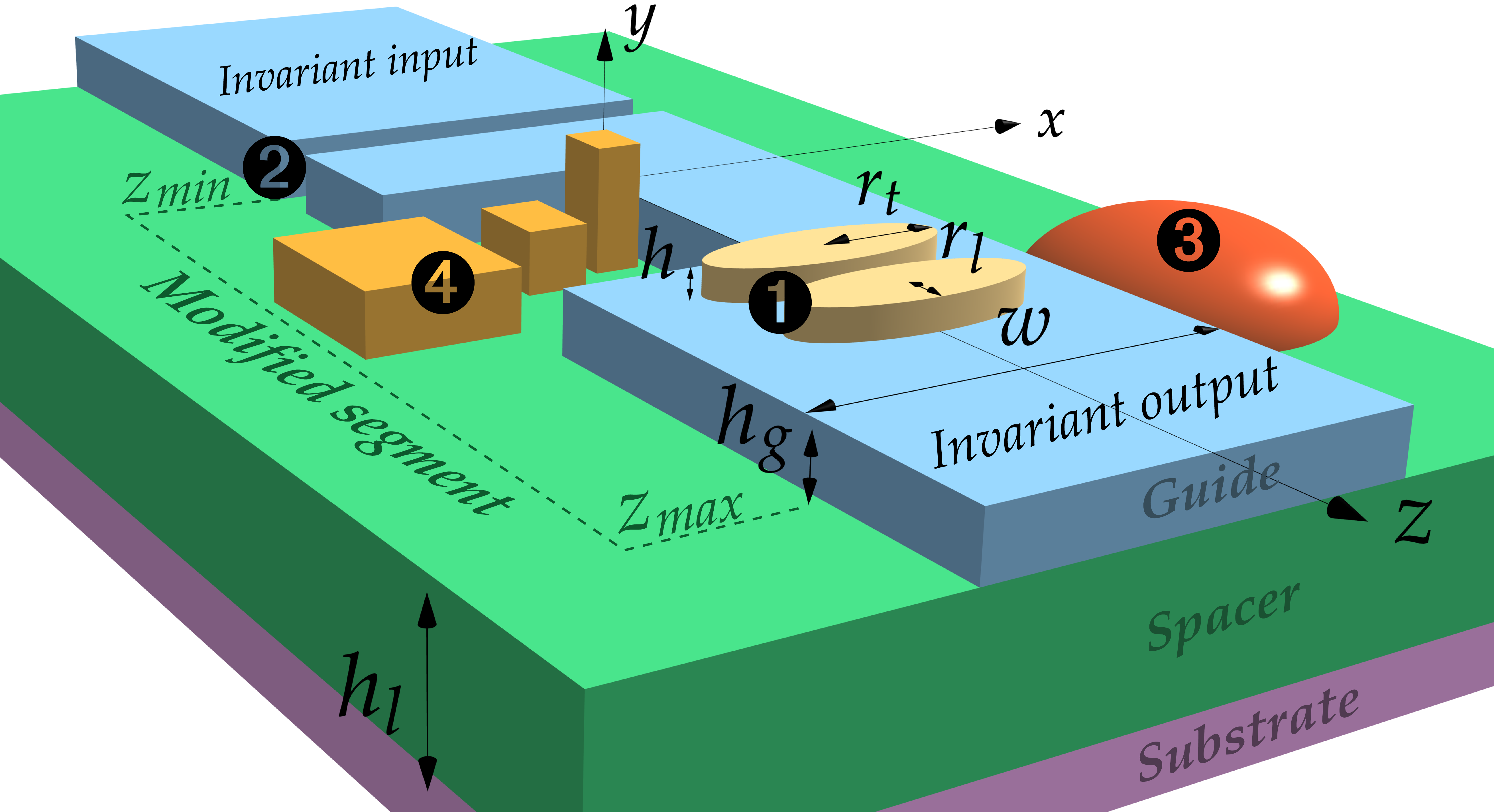}
  \caption{Scheme of the $z$-invariant structure (substrate in purple, low index layer in green 
   and a rectangular waveguide in blue) with various discontinuities (or obstacles) breaking the $z$-invariance locally in a region called ``modified segment''. Discontinuities can be ellipsoidal patches above the guiding layer labelled~\ding{202}, holes~\ding{203} in the guiding layer, obstacles or resonators next to the waveguide~\ding{204} or even a combination of all~\ding{205}.
  }
  \label{fig:scheme}
\end{figure}

In this section, a direct -- as opposed to modal -- scattering approach is introduced. A typical and realistic structure is sketched in Fig.~\ref{fig:scheme}: A $z$-invariant dielectric rectangular waveguide (of width $w$ and thickness $h_g$ in blue) is deposited on a low index spacer (of thickness $h_l$, in green) lying on a semi-infinite substrate (in purple). The $z$-invariance of this guiding structure is locally broken, by adjunction of a finite number of obstacles. These obstacles can be in practice any bounded modification of permittivity: Ellipsoidal patches above the guiding layer labelled~\ding{202} in Fig.~\ref{fig:scheme}, holes~\ding{203} in the guiding layer, obstacles or resonators next to the waveguide~\ding{204} or even a combination of all~\ding{205}\dots Note that the method applies irrespectively of the number of layers of the $z$-invariant structure and that the obstacles can be arbitrarily shaped and located in (or above) the structure. It is shown how the obstacles  (more generally the modified waveguide segment) perturb a mode propagating in the $z$-invariant structure. A first step consists in the numerical computation of the modes of the invariant structure, which are used in a second step as incident fields for the full 3D structure.

\subsection{Obtaining  the incident fields}\label{sec:3Ddirect1}
The classical guiding $z$-invariant structure is characterized by its permittivity function defined by parts as: 
\begin{equation}
  \epsrdd(x,y)=
  \left \{
  \begin{array}{l}
    \epsrin{g} \mbox{ in the guide,} \\
    \epsrin{l} \mbox{ in the low index region,} \\
    \epsrin{s} \mbox{ in the substrate,} \\
    \epsrin{t} \mbox{ in the superstrate}
  \end{array} \right. .
\end{equation}

Between $z_{min}$ and $z_{max}$, one can now break the $z$-invariance by a local modification of the permittivity function which leads to
a 3D scattering problem, which in turn can be characterized by its permittivity function defined by parts:
\begin{equation}
  \epsrddd(x,y,z)=
  \left \{
  \begin{array}{l}
    \epsrin{g} \mbox{ in the guide,} \\
    \epsrin{l} \mbox{ in the low index region,} \\
    \epsrin{s} \mbox{ in the substrate,} \\
    \epsrin{t} \mbox{ in the superstrate},\\
    \epsrin{d}(x,y,z) {\crevme\mbox{ in the obstacles }}\\
        \hspace{.5cm} {\crevme\mbox{ of the modified segment}}
  \end{array} \right. .
\end{equation}

The starting point consists in computing the modes of an annex problem formed by the $z$-invariant 
structure solely. This is a very classical problem~\cite{snyder1983optical,renversez2012foundations} where one introduces the ansatz 
$\bE(\bx)=\be(\x)e^{-i(\om_0 t-\beta z)}$ in the source-free Helmholtz equation:
\begin{equation}\label{eq:Hdd}
  \curl\left(\tensmurdd^{-1}\,\curl\bE\right) = \tensepsrdd\left(\frac{\om_0}{c}\right)^2 \bE
\end{equation}
{\crevi for a given real angular frequency $\om_0$}. Note that the relative permittivity and permeability in \eqref{eq:Hdd} are tensors fields since cartesian PMLs adapted to each domain with infinite extension (superstrate, low index region and substrate) are used to damp the radially blowing leaky modes of this open structure~\cite{renversez2012foundations}. The domains $\Omega$ involved in all the formulations of the paper are correspond to geometrical domains surrounded by appropriate PMLs of finite thicknesses. 

It results in a quadratic non-Hermitian eigenvalue problem amounting to find non trivial pairs $(\beta_k,\bek)\in\mathbb{C}\times H^1(\Omega_{2D},\mathbf{curl})$ such that : 
\begin{equation}\label{eq:devstrongform2}
  \begin{split}
    &\curl\left(\tensmurdd^{-1}\,\curl \bek\right)-k_0^2\,\tensepsrdd\,\bek \\
    +i\betak&\left[\bzh\times \left(\tensmurdd^{-1}\curl \bek\right) + \curl\left(\tensmurdd^{-1}\,\bzh\times\bek \right)\right] \\
    +(i\betak)^2\,& \bzh\times\left(\tensmurdd^{-1}\,\bzh\times\bek \right)=\mathbf{0}\,,
  \end{split}
\end{equation} 
with $ k_0 := \om_0/c$.
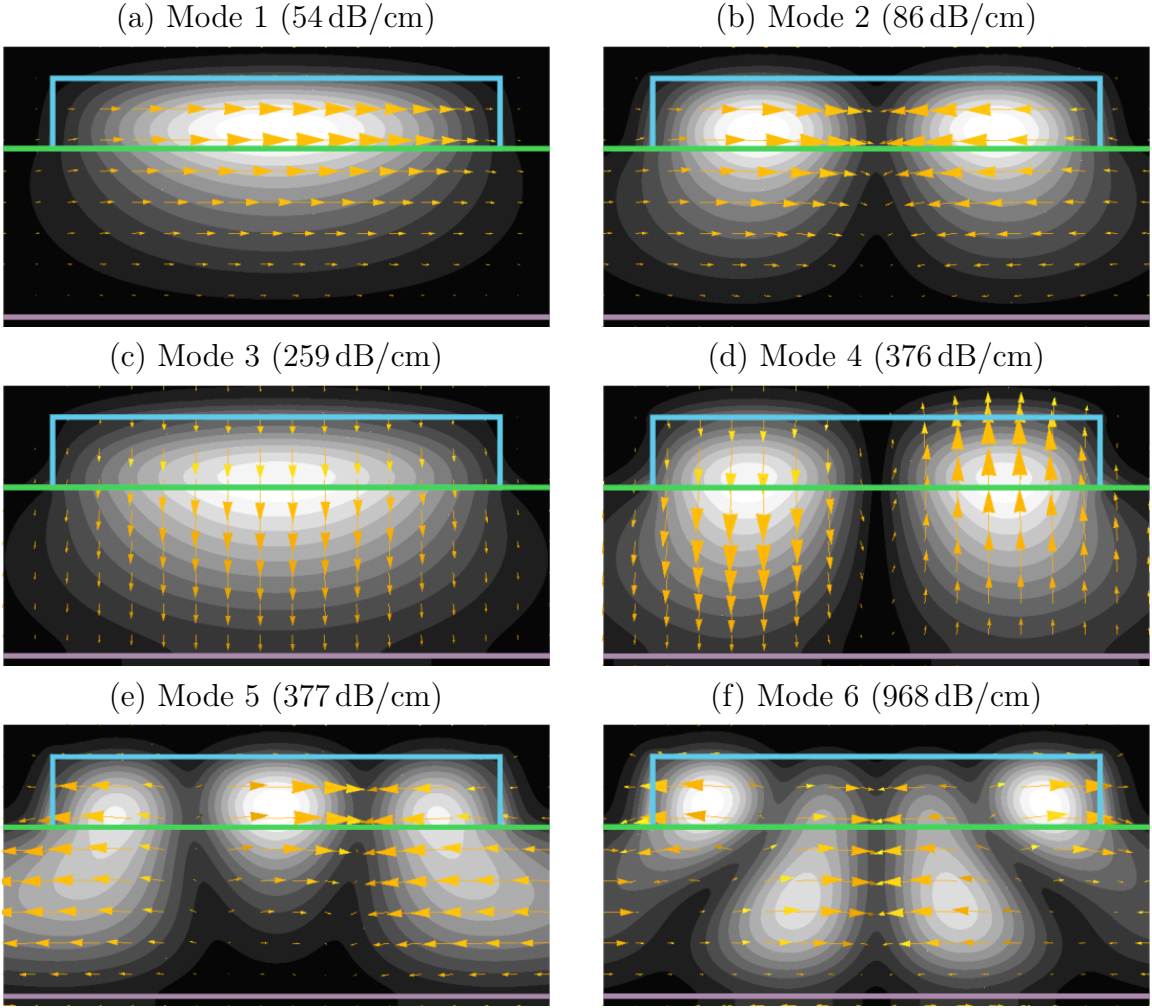
\begin{figure}[h]
  \centering
  \begin{tikzpicture}
    \node[inner sep=0pt] (modeone) at (-.4,9.0) {\includegraphics[width=7.2cm]{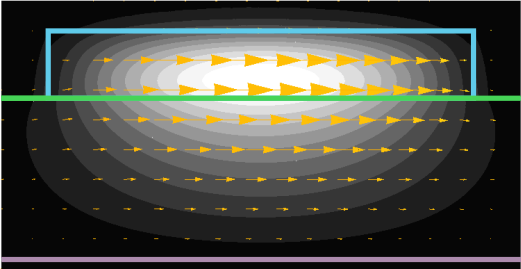}};
    \node[inner sep=0pt] (modetwo) at (7.5,9.0) {\includegraphics[width=7.2cm]{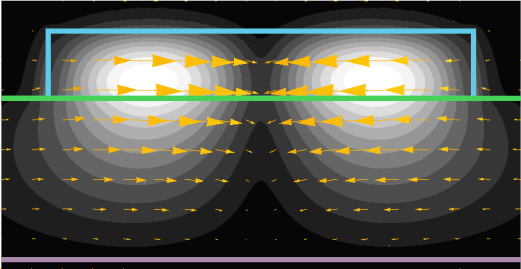}};
    \node[inner sep=0pt] (modethr) at (-.4,4.5) {\includegraphics[width=7.2cm]{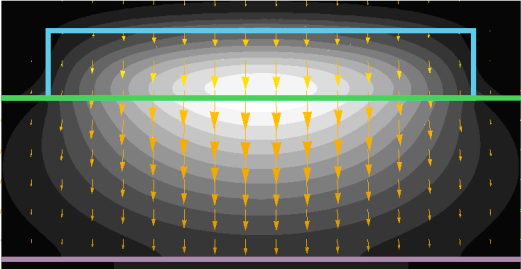}};
    \node[inner sep=0pt] (modefou) at (7.5,4.5) {\includegraphics[width=7.2cm]{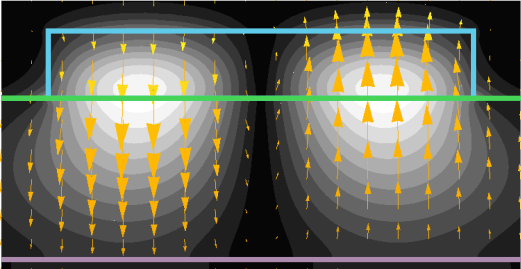}};
    \node[inner sep=0pt] (modefiv) at (-.4,0.0) {\includegraphics[width=7.2cm]{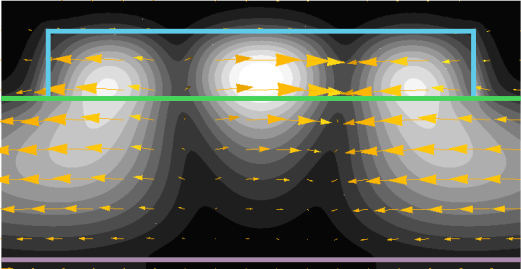}};
    \node[inner sep=0pt] (modesix) at (7.5,0.0) {\includegraphics[width=7.2cm]{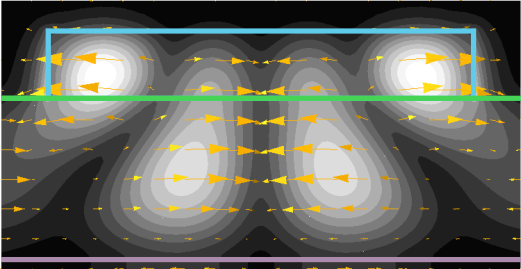}};
    \node[above] at (modeone.north) {(a) Mode 1 (\SI{ 54}{dB/cm})};
    \node[above] at (modetwo.north) {(b) Mode 2 (\SI{ 86}{dB/cm})};
    \node[above] at (modethr.north) {(c) Mode 3 (\SI{259}{dB/cm})};
    \node[above] at (modefou.north) {(d) Mode 4 (\SI{376}{dB/cm})};
    \node[above] at (modefiv.north) {(e) Mode 5 (\SI{377}{dB/cm})};
    \node[above] at (modesix.north) {(f) Mode 6 (\SI{968}{dB/cm})};
  \end{tikzpicture}
  \caption{{\crevii The six modes with smallest attenuation supported by the $z$-invariant structure at $\lambda_0=\mic{7.7}$. The edges of the cross-section are represented in colors matching the domains shown in Fig.~\ref{fig:scheme}. The eigenvalue corresponding to mode~1 in the inset~(a) has the smallest imaginary part. The power attenuation (defined \cite{renversez2012foundations} in dB/cm as -$2000\,\mathrm{Im}\{\beta_k\}/\mathrm{ln}(10)$) is given at the top of each inset. The black and white maps (white is high) represent the norm of the electric eigenfields $|\bek|$ in the waveguide cross-section. The orange arrows indicate the real part of the electric eigenfields $\bek$.}}
  \label{fig:modes2D}
\end{figure}

This equation can be solved using a mixed finite element formulation involving edge elements for the discretization of the transverse component $(e_x,e_y)$ coupled  to a nodal basis for the (continuous) longitudinal component $e_z$. {\crevi The rather lengthy details of the resulting weak formulation can be found in Refs.~\cite{nicolet2004modelling,renversez2012foundations}}.

%

Throughout the paper, the following numerical values are considered for the $z$-invariant waveguide~\cite{kuriakose17Measurement-ultrafast-optical-Kerr-effect-chalco}:
The operating freespace wavelength 
$\lambda_0=\mic{7.7}$, 
$\epsrin{g}=7.1824$ ($\mbox{Se}_4$),
$\epsrin{l}=6.2001$ ($\mbox{Se}_2$),
$\epsrin{s}=11.69024481$ (silicon) \cite{chandler2005high},
$\epsrin{t}=1$ (air),
$w_g=\mic{14}$, 
$h_g=\mic{2.2}$
and $h_l=\mic{5.3}$. All the materials are assumed to be non-magnetic: $\tensmurdd=\tensid$ (except in the PMLs where $\tensmurdd$ takes the appropriate value), where $\tensid$ is the $3\times3$ identity tensor.

{\crevii The modes of this structure associated with eigenvalues with lowest imaginary parts are depicted in Fig.~\ref{fig:modes2D}(a-f), sorted in ascending attenuation ({\textit i.e.} $\mbox{Im}\{\beta_1\}$ is the smallest). The black and white colormaps show the norm of the electric eigenfields $|\be_k|$ in Fig.~\ref{fig:modes2D}(a-f) and the orange arrows represent the real part of $\be_k$, allowing to distinguish a TE-like mode from a TM-like one. These six modes are also those with electric eigenfield most confined into the core region of the structure (ridge).}
  
Finally, all geometries and conformal meshes have been obtained using the  Gmsh software~\cite{gmsh} and all the finite element formulations in this article are implemented thanks to the flexibility of the finite element software GetDP~\cite{getdp}. Open source models allowing to retrieve most of the results of this article are provided \cite{code}.

\subsection{Computation of the scattered field}\label{sec:3Ddirect2}
One can now use any of these 2D modes $\bEdd{k}:=\be_k\,e^{i(\betak z-\omega_0 t)}$ as an incident field $\bEinc$ on the obstacles and look for $\bEtot$, the total field solution of the source-free Helmholtz equation:
\begin{equation}\label{eq:Hddd}
  -\curl\left[\tensmurddd^{-1}\,\curl \bEtot \right] +  k_0^2\,\tensepsrddd\,\bEtot = \mathbf{0}.
\end{equation}
Let us define the scattered field as $\bEd \equiv \bEtot-\bEinc$ and from the linearity of Eqs.~(\ref{eq:Hdd},\ref{eq:Hddd}), we obtain the following scattering problem:
\begin{equation}\label{eq:Hdddrad}
  -\curl\left[\tensmurddd^{-1}\,\curl \bEd \right] + k_0^2\,\tensepsrddd\,\bEd = k_0^2\,(\tensepsrdd-\tensepsrddd)\bEinc.
\end{equation}
Note that the support of the effective sources $(\tensepsrdd-\tensepsrddd)$ in this scattering problem has to be bounded to ensure a proper outgoing wave condition \cite{zolla1996method} to the scattered field $\bEd$, which is the case in our examples. Finally, 3D cartesian PMLs are used to bound the computational domain \cite{renversez2012foundations,berenger94perfec-match-layer} {\crevii as shown in grey lines in Fig.~\ref{fig:poynting}}. Compared to a total field approach with a port condition \cite{jin02FEM-electromag}, it is stressed that the electromagnetic sources of our equivalent radiation problem are located within the discontinuities. The PMLs of elongated structures are naturally built to damp fields radiating from the center of the computational box more efficiently than the total field radiating from a port located at one extremity of the elongated box, the resulting total field being more grazing than the scattered field when entering the PMLs.

\subsection{Energy balance}\label{sec:3Ddirect3}
\begin{figure}
  \centering
  \includegraphics[width=.7\columnwidth]{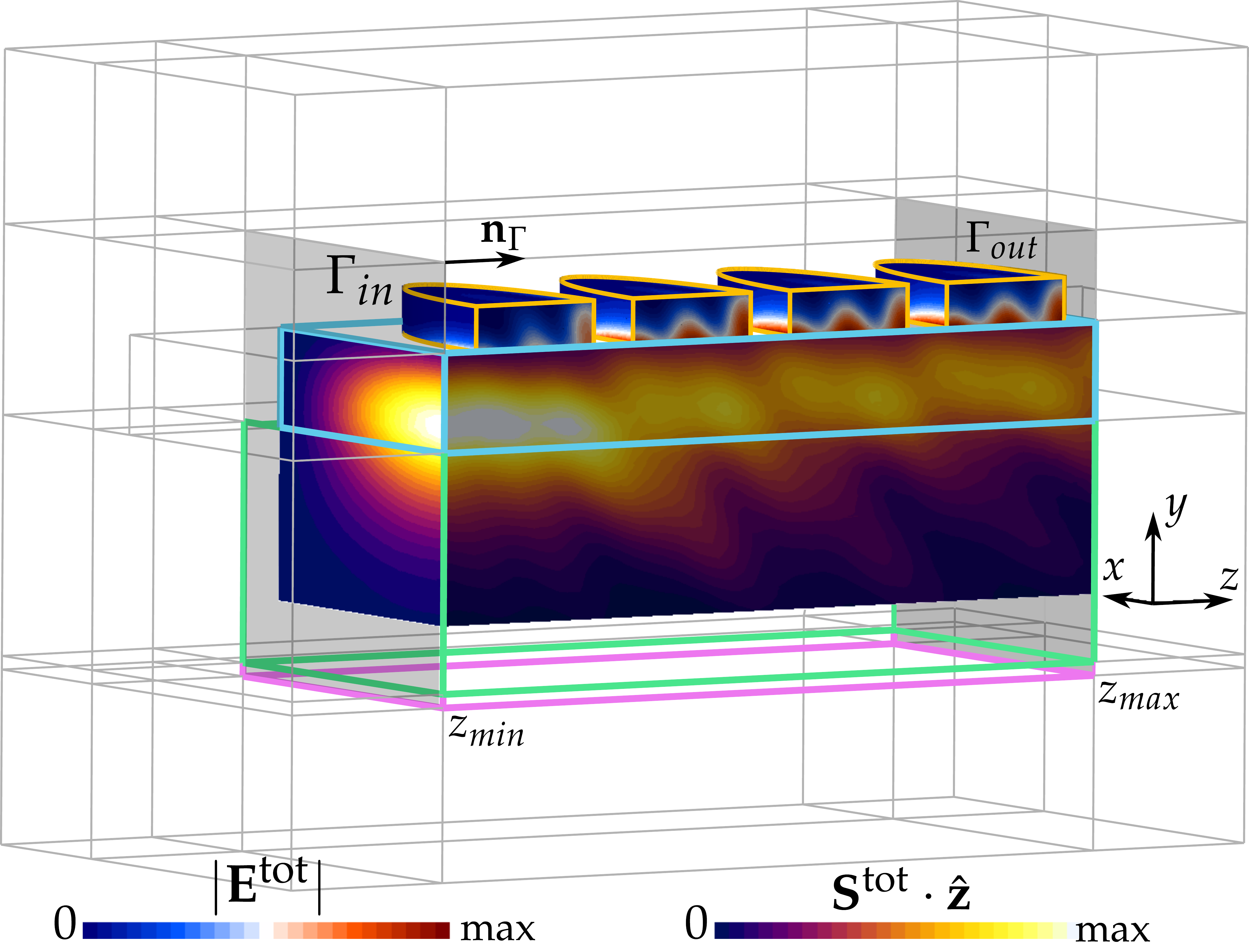}
  \caption{
      Cuts of the {\crevii $z$-component of the total Poynting vector ($\bStot\cdot\bzh$ in purple/yellow colors) and norm of the total field ($|\bEtot|$ in blue/red colors) inside the four lossy obstacles (ellipsoidal patches) above the waveguide. The computational domain represents half of the structure due to the symmetry properties of both the geometry and the incident field. The colored edges represent the actual geometry of the structure with the same color code as in Fig.~\ref{fig:scheme} and the grey ones the cartesian PMLs adapted to each physical domain.} }
  \label{fig:poynting}
\end{figure}

The Poynting vectors associated with the incident, diffracted and total fields are classically defined by respectively $\bSi=\real{\bEinc\times\overline{{\bHinc}}}/2$, $\bSd=\real{\bEd\times\overline{\bHd}}/2$ and $\bStot=\real{\bEtot\times\overline{\bHtot}}/2$, where the horizontal bar means complex conjugation. Then, the incoming, transmitted, reflected and absorbed powers
can be defined as respectively :
\begin{equation}\label{eq:power}
  \begin{split}
    P_{in}  &= \int_{\Gamma_{in}}\bSi\cdot \bn_\Gamma\,\mathrm{d}S,\\
    P_{tr}  &= \int_{\Gamma_{out}}\bStot\cdot \bn_\Gamma\,\mathrm{d}S,\\
    P_{ref} &= \int_{\Gamma_{in}}\bS^{d}\cdot \bn_\Gamma\,\mathrm{d}S \mbox{ and }\\
    P_{abs} &= \frac{\varepsilon_0\,\omega_0}{2}\int_{\Omega_{d}} {\crevme \mathrm{Im}\{\varepsilon_{r,d}\}}\,|\bEtot|^2\,\mathrm{d}\Omega,
  \end{split}
\end{equation}
where $\Gamma_{in}$ and $\Gamma_{out}$ are {\crevii the transverse plane surfaces before and after the obstacles depicted in transparent grey color in Fig.~\ref{fig:poynting}}, $ \bn_\Gamma$ is the unit vector normal to $\Gamma_{in}$ and  $\Gamma_{out}$ and $\Omega_{d}$ is the support of the diffractive obstacles or of the localized region where the waveguide opto-geometrical parameters are modified.
Finally one can define transmission (T), reflection (R) and absorption (A) coefficients as :
\begin{equation}\label{eq:RTA}
  T=\frac{P_{tr}}{P_{in}} \mbox{ , } R=-\frac{P_{ref}}{P_{in}}\mbox{ and } A=\frac{P_{abs}}{P_{in}}.
\end{equation}

For clarity, Fig.~\ref{fig:poynting} illustrates the quantities at stake in the energy balance. This numerical set up is obtained for an incident field set to $\bEdd{1}$ (cf Fig.~\ref{fig:modes2D}(a)) with four ellipsoidal lossy patches placed above the same waveguide as in Section~\ref{sec:3Ddirect1}. This particular configuration will be discussed in detail in Section~\ref{sec:disc}. {\crevii The red/blue colormap represents the norm of the total electric field $|\bEtot|$ involved in the computation of the Joule losses, \emph{i.e.} $P_{abs}$ in \eqref{eq:power}. The purple/yellow colormap represents three cuts of the $z$~component of the total Poynting vector $\bStot$ on three selected plane surfaces. The first cut is taken at $z=z_{min}$ (see left side of the figure), another one at $z=z_{max}$ (right side of the figure) and the last one along the symmetry plane $x=0$ of the structure. In this last cut, it is clear that the perturbation induced by the objects affects the $z$~component of the total Poynting vector since in absence of scattering objects above the ridge, this map would be constant along $z$}.

{\crevi In this example, the transmission $T$ reaches 0.688, the reflection $R=0.007$ and the absorption $A=0.224$  ($T+R+A=0.904$). Note that $R$, $T$ and $A$ are defined here in order to match commonly measured quantities does but do not add up to unity. It is nonetheless expected since their sum does not correspond to a full Poynting balance. It can be easily completed by adding the flux contributions from the surfaces parallel to the $zOx$ and $zOy$ planes, which represent the extra radiative leakage induced by the modified segment ($9.6\%$ in the present case). 
     
As shown in Sec.~\ref{sec:3Ddirect4}, the scattering process can be further precised by expansion of the diffracted and total fields outside the modified segment on the modes of the 2D invariant structure.
}

\subsection{\crevi Discretization and convergence}

{\crevi In the 2D eigenvalue problem of Sec.~\ref{sec:3Ddirect1}, the longitudinal component of the electric field, which is a continuous scalar field, is discretized using classical $P_2$ nodal elements having one Degree Of Freedom (DOF) per node and one DOF per edge. The transverse components are discretized using edge elements of the second order (2 degrees of freedom per edge). For eigenvalue problems, the GetDP software relies on the high performance library SLEPc \cite{Hernandez2005-SSF} which implements advanced Krylov subspace methods for computing a small amount of eigenvalues of the large sparse matrices.

The 3D scattering problem uses high order Webb hierarchical edge elements \cite{geuzaine1999convergence,webb1993hierarchal,jin02FEM-electromag} with 26 DOFs per tetrahedron (3 DOFs per edge, 2 DOFs per face). The direct problem described in section~\ref{sec:3Ddirect2} is solved using the direct solver MUMPS \cite{mumps-userguide} interfaced in GetDP}. 

{\crevii The convergence of the absorption $A$ as a function of the mesh refinement is shown in Fig.~\ref{fig:conv}(a). The mesh size is parametrized by $n$ (in abscissa) and decreases as $\lambda_0/(n\,\mathrm{Re}\{\sqrt{\varepsilon_r}\})$ as $n$ increases. In other words, $n$ represents the average number of tetrahedrons per wavelength inside a dielectric material of relative permittivity $\varepsilon_r$. Note that in metals, the relevant physical length to consider for a proper spatial sampling of the field would be the skin depth rather that the wavelength. For $n$=1, that is roughly one tetrahedron per wavelength, the computational box in Fig.~\ref{fig:poynting} leads to 32000~DOFs solved in \SI{3}{\second} on a laptop equipped with 4~cores 16~Gb of RAM memory. In this case, the local values of the field are poorly approximated, but the order of magnitude of integral quantities such as the absorption is relevant, as can be noticed on the left side of the convergence plot shown in Fig.~\ref{fig:conv}(a). For $n$=4, the number of DOFs is 650000, the model still runs on the same laptop within \SI{4}{\minute}. For $n$=7, the number of DOFs becomes about 3 millions and a workstation equipped with 24~cores and 256~Gb RAM memory was used for a runtime of \SI{30}{\minute}. Five significant digits are then obtained on energy-related quantities such as the absorption.} 


\subsection{Modal expansion of the scattered field}\label{sec:3Ddirect4}
The modes of the 2D invariant structure satisfy the following bi-orthogonality 
condition equivalent to the one given in Ref.~\cite{sammut1976leaky,snyder1983optical} which provides the normalization of each leaky modes:
\begin{equation}\label{eq:orth}
 \int_S \be_j\times\bh_k\cdot\bzh\,\mathrm{d}S = \int_S \be_k\times\bh_j\cdot\bzh\,\mathrm{d}S=A_k\delta_{kj} \; \mbox{, where} 
\end{equation}
$S$ is an infinite cross-section of the open waveguide. In the case of leaky modes, it is suggested in~\cite{sammut1976leaky} to perform a complex change of space variable as one moves far away from the waveguide to damp the exponential growth of the leaky mode. In our finite element approach that includes the PMLs which are an analytical continuation of the space variables, this integration simply corresponds to integrate over a full cross-section of computational domain including the PML regions. Therefore it is stressed that the cross-sections considered hereafter include the PMLs.
Hence, away from the obstacles, it is possible to expand the scattered field as $\bEd=\sum_k r_k \bEdd{k}$ and the total field as $\bEtot=\sum_k t_k \bEdd{k}$ where the reflection and transmission coefficients are simply given as:
\begin{equation}
  \label{eq:rntn}
  \left \{
  \begin{array}{l}
    t_k=\dint_{S_{out}} \bEtot\times\bh_k\cdot\bzh\,\mathrm{d}S / A_k  \\[4mm]
    r_k=\dint_{S_{in}} \bEd\times  \bh_k\cdot\bzh\,\mathrm{d}S / A_k
  \end{array} \right. 
\end{equation}
where $S_{in}$ can be any transverse section before the obstacles ($z<z_{min}$) and $S_{out}$ can be any transverse section after the obstacles ($z>z_{out}$). Note that this formalism can be extended to the computation of the scattering matrix of the waveguide, by considering sequentially all the modes of the invariant structure computed in Sec.~\ref{sec:3Ddirect1} as  incident fields. In this example, the modulus of the mean value of the off-diagonal coefficients of the $12\times12$ bi-orthogonality matrix (see \eqref{eq:orth}) is less than $10^{-3}$ smaller than the modulus of the mean value of the diagonal coefficients $A_k$.

\begin{figure}[t] 
  \centering
  \includegraphics[width=.9\linewidth]{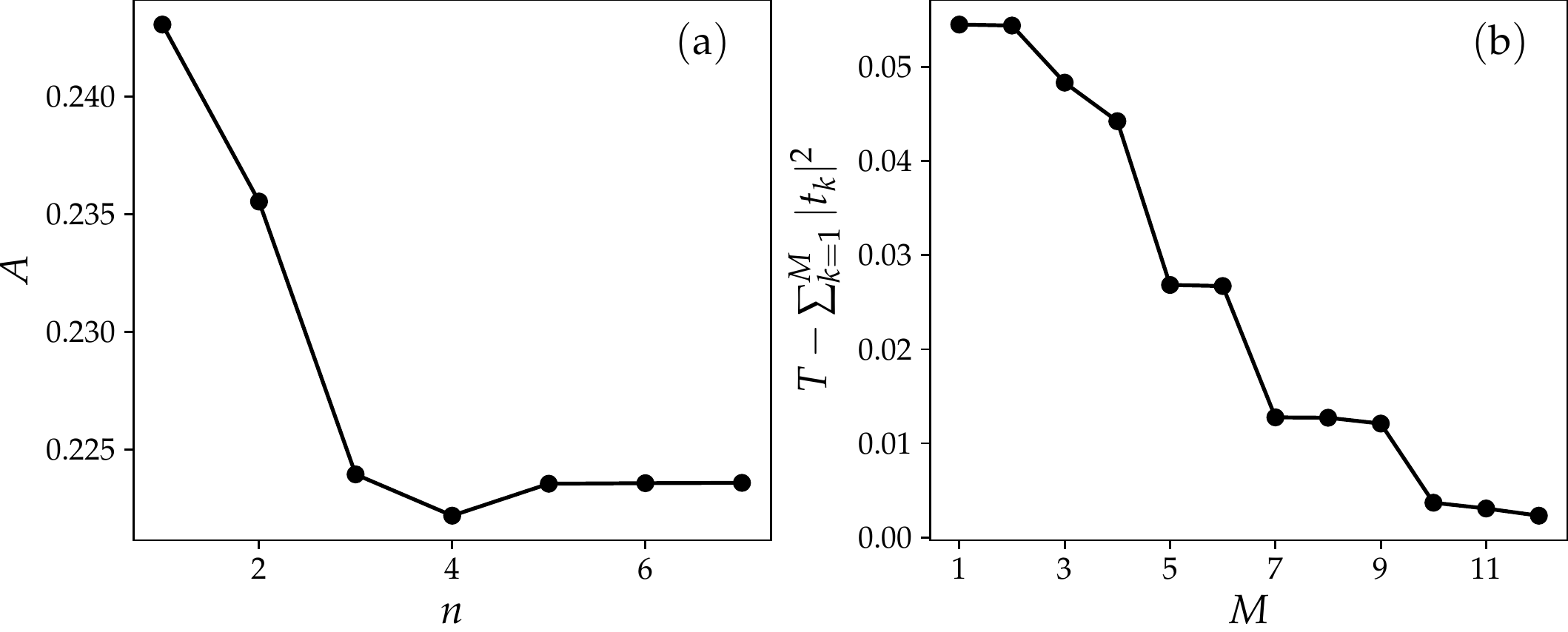}
  \caption{(a) {\crevii Convergence of the absorption $A$ of the lossy ellipsoidal patches as a function of the mesh size decreasing as $\lambda_0/(n\,\mathrm{Re}\{\sqrt{\varepsilon_r}\})$ as $n$ (in abscissa) increases. (b) The quantity $T-\sum_{k=1}^M |t_k|^2$ as a function of the truncation order $M$ for $n$=7.}}
  \label{fig:conv}
\end{figure}

{\crevi The total field within a leaky guiding structure can be expanded \cite{snyder1983optical} as discrete sum over bounded modes plus an integral over the continuous spectrum. When using finite size PMLs, the continuous spectrum becomes discrete and the integral contribution turns into a discrete sum. Besides, the power orthogonality between the modes holding for self-adjoint eigenvalue problems such as the perfect metallic waveguide now fails in the our non-Hermitian case. This can be seen in the bi-orthogonality relations (see \eqref{eq:orth}) that involve the magnetic field rather than its complex conjugate. However, we are in a weakly leaky regime \cite{sammut1976leaky} were the weakly leaky modes decay rapidly as the radial distance $r=\sqrt{x^2+y^2}$ increases until experiencing an exponential blow at even larger radial distances. In this regime, it is interesting to see that the power exchange between the most highly confined modes (associated with eigenvalues with small imaginary part compared to their real part) can be neglected. In short, $\sum_k |t_k|^2\rightarrow T$ and $\sum_k |r_k|^2\rightarrow R$ hold to a very good approximation. In the particular configuration described in Fig.~\ref{fig:poynting}, one obtains the values $\sum_{k=1}^{12}|t_k|^2=0.685$ (with a major contribution from $|t_1|^2=0.632$, to be compared to $T$=0.688 obtained in Sec.~\ref{sec:3Ddirect3}) and $\sum_{k=1}^{12}|r_k|^2=0.0007$. The comparison between $T$ and $\sum_{k=1}^M |t_k|^2$ as a function of the truncation order $M$ is shown in Fig.~\ref{fig:conv}(b) for a fine mesh with $n=7$.}

\section{Modes of the infinitely periodic 3D structure}\label{sec:3Dmodal}
\subsection{Variational formulation of the spectral problem}\label{sec:3Dmodal1}
In this section, we are now interested in a 3D spectral problem with one direction of periodicity defined by \begin{equation}\label{eq:L3D}
  \curl\left[\tensmurp(\bx)^{-1}\curl \bE\right]=\tensepsrp(\bx,\omega_0)\left(\frac{\omega_0}{c}\right)^2\,\bE\,.
\end{equation}
where $ \tensepsrp(\bx,\omega_0)$ and  $\tensmurp(\bx)$ are respectively the permittivity and  permeability tensor fields at a fixed frequency $\om_0$ exhibiting a 1D $d$-periodicity along $Oz$. Bloch's theorem states that, without loss of generality, one can look for solutions for the electric field $\bE$ under the form~\cite{joannopoulos95photon-crist-modlin-flow-light,renversez2012foundations} :
\begin{equation}\label{eq:solform}
  \bE=\bEp(x,y,z)\,e^{-i(\omega_0 t - \gamma z)}\,,
\end{equation}
where $\bE_\#$ is a $d$-periodic function in $z$ and $\gamma$ is the Bloch variable lying in the first reduced Brillouin zone [0,$\pi/d$].

One can choose to set $\gamma$ to a real value lying in the first Brillouin zone and  to look for ($\omega_{\gamma,i}$,$\bE_{\gamma,i}$) eigenvalues and eigenvectors,  by imposing Bloch conditions on the z-transverse surfaces of the cell and making the use of \eqref{eq:L3D}. An alternative option amounts to set $\omega_0$ to a real value, inject the ansatz in \eqref{eq:solform} into \eqref{eq:L3D} and look for eigenvectors under the form of the periodic part $\bEp$ of the Bloch wave along with corresponding eigenvalue $\gamma$. In this latter case, two equations are to be fulfilled:
\begin{subequations} 
  \begin{empheq}[left=\empheqlbrace]{align}
    \label{eq:modal3Dcurl}-\curl\left[\tensmurp^{-1}\curl\left[\bEp e^{i\gamma z}\right]\right]+k_0^2\,\tensepsrp(\bx)\,\bEp e^{i\gamma z}&=0\\
    \label{eq:modal3Ddiv}\div\left[\tensepsrp(\bx)\,\bEp e^{i\gamma z}\right]&=0
  \end{empheq}
  .
\end{subequations}

It is stressed that the invariant 2D problem described in Sec.~\ref{sec:3Ddirect1} is a particular case of this 3D problem, an invariant structure along $z$ being trivially periodic in $z$ with arbitrary period. Unsurprisingly, expanding the $\mathbf{curl}$ term in \eqref{eq:modal3Dcurl} in order to get rid of the  $e^{i\gamma z}$ dependency, leads to an expression similar to the $z$-invariant counterpart of the  problem (see \eqref{eq:devstrongform2}):
 
\begin{equation}\label{eq:devstrongform3}
  \begin{split}
    -\curl\left[\tensmurp^{-1}\curl \bEp\right]&+k_0^2\,\tensepsrp(\bx)\,\bEp \\
    -i\gamma\, & \bzh\times \left[\tensmurp^{-1}\curl \bEp\right] \\
    -i\gamma\, & \curl\left(\tensmurp^{-1}\,\bzh\times\bEp \right) \\
    -(i\gamma)^2\,& \bzh\times\left(\tensmurp^{-1}\,\bzh\times\bEp \right) =\mathbf{0} \,.\\
  \end{split}
\end{equation}
In a variational way, after classically integrating by part two curl operators, it holds that for any $\mathbf{W}\in H_\#^1(\Omega,\curl)$ :
\begin{equation}\label{eq:devstrongform3}
  \begin{split}
    -&\dint_\Omega \left[\tensmurp^{-1}\curl \bEp\right]\cdot\curl \bW \d\Omega \\
    +&\dint_\Omega k_0^2\,\tensepsrp(\bx)\,\bEp \cdot\bW \d\Omega\\
    -i\gamma\, &\dint_\Omega  \bzh\times \left[\tensmurp^{-1}\curl \bE\right] \cdot\bW \d\Omega\\
    -i\gamma\, &\dint_\Omega  \left(\tensmurp^{-1}\,\bzh\times\bEp \right) \cdot\curl\bW \d\Omega \\
    +(i\gamma)^2\,&\dint_\Omega  \left(\tensmurp^{-1}\,\bzh\times\bEp \right)\cdot(\bzh\times\bW) \d\Omega\\
    - &\dint_{\partial\Omega} \left[\bn_{|\partial\Omega}\times\left(\tensmurp^{-1}\curl \bEp\right)\right]\cdot \bW \d S\\
    - i\gamma&\dint_{\partial\Omega} \left[\bn_{|\partial\Omega}\times\left(\tensmurp^{-1}\,\bzh\times\bEp \right) \right]\cdot \bW \d S \\ 
  & =0
\end{split}
\end{equation}
Note that the two boundary terms recombine into $-\int_{\partial\Omega} [\bn_{|\partial\Omega}\times(\tensmurp^{-1}(\curl \bEp+i\gamma\bzh\times\bEp) )]\cdot \bW \d S\propto\int_{\partial\Omega} [\bn_{|\partial\Omega}\times\bH]\cdot \bW \d S$ so that setting a Dirichlet or Neumann natural condition for $\bEp$ on non-periodic faces of the domain ({\it i.e}. the PML bounds) actually corresponds to a Dirichlet or Neumann natural condition for $\bH$.

The divergence condition in \eqref{eq:modal3Ddiv} has to be handled carefully. Indeed, we are looking for divergence free solutions such that $\div(\tensepsrp\,\bE)=0$, that is:
\begin{equation}\label{eq:div}
  \div\left(\tensepsrp\,\bEp e^{i\gamma z}\right)=0=\div(\tensepsrp\,\bEp)+i\gamma\bzh\cdot(\tensepsrp\,\bEp)\\
\end{equation}
Consequently, $\tensepsrp\,\bEp$ is not divergence-free and, from the variational point of view, the following holds 
for any $\varphi\in H_\#^1(\Omega)$:
\begin{equation}\label{eq:div}
  \begin{split}
    \int_\Omega&\left[\div\left(\tensepsrp\,\bEp\right)+i\gamma\bzh\cdot\left(\tensepsrp\,\bEp\right)\right]\,\overline{\varphi}\,\d\Omega=0\\
    =-\int_\Omega&\tensepsrp\,\bEp\cdot\,\overline{\grad\varphi}\,\d\Omega+i\gamma\int_\Omega\bzh\cdot\left(\tensepsrp\,\bEp\right)\,\overline{\varphi}\,\d\Omega\,,
  \end{split}
\end{equation}
{\crevme where the boundary term arising from the integration by part vanishes due to periodicity and homogeneous conditions at the back of the PMLs.}

Finally, the proper way to ensure the divergence condition \cite{lackner2019} in a weak sense is to use $\varphi$ as a Lagrange multiplier. We are now in position to reformulate the eigenvalue problem at stake in this section. We are looking for non trivial pairs $\gamma_k,(\bEpk,\varphi_k)\in \mathbb{C}\times(H^1_\#(\Omega,\mathrm{curl})\times H_\#^1(\Omega))$ such that:

\begin{subequations} \label{eq:weakmodal3D}
  \begin{empheq}[left=\empheqlbrace]{align}
    \label{eq:weakmodal3Dcurl}
    \begin{split}
      -&\dint_\Omega \tensmurp^{-1}\curl \bEpk \cdot\curl \bW \d\Omega\\
      +&\dint_\Omega k_0^2\,\tensepsrp(\bx)\,\bEpk \cdot\bW \d\Omega\\
      -i\gamma &\dint_\Omega  \bzh\times \left(\tensmurp^{-1}\curl \bE\right) \cdot\bW \d\Omega\\
      -i\gamma &\dint_\Omega  \left(\tensmurp^{-1}\,\bzh\times\bEpk \right) \cdot\curl\bW \d\Omega \\
      +(i\gamma)^2&\dint_\Omega  \left(\tensmurp^{-1}\,\bzh\times\bEpk \right)\cdot(\bzh\times\bW) \d\Omega\\
      +&\dint_\Omega  \tensepsrp\grad\varphi_k\cdot\bW \d\Omega\\
      +i\gamma &\int_\Omega\tensepsrp\varphi_k\bzh\cdot\bW\d\Omega=0 
    \end{split}\\[2mm]
    \label{eq:weakmodal3Ddiv}
    \begin{split}
       &\int_\Omega\tensepsrp\,\bEpk\cdot\,\overline{\grad\varphi_k}\,\d\Omega\\
       -i\gamma&\int_\Omega\bzh\cdot\left(\tensepsrp\,\bEpk\right)\,\overline{\varphi_k}\,\d\Omega=0 
    \end{split}
  \end{empheq}
\end{subequations}
\begin{figure}
  \centering
  \includegraphics[width=.65\columnwidth]{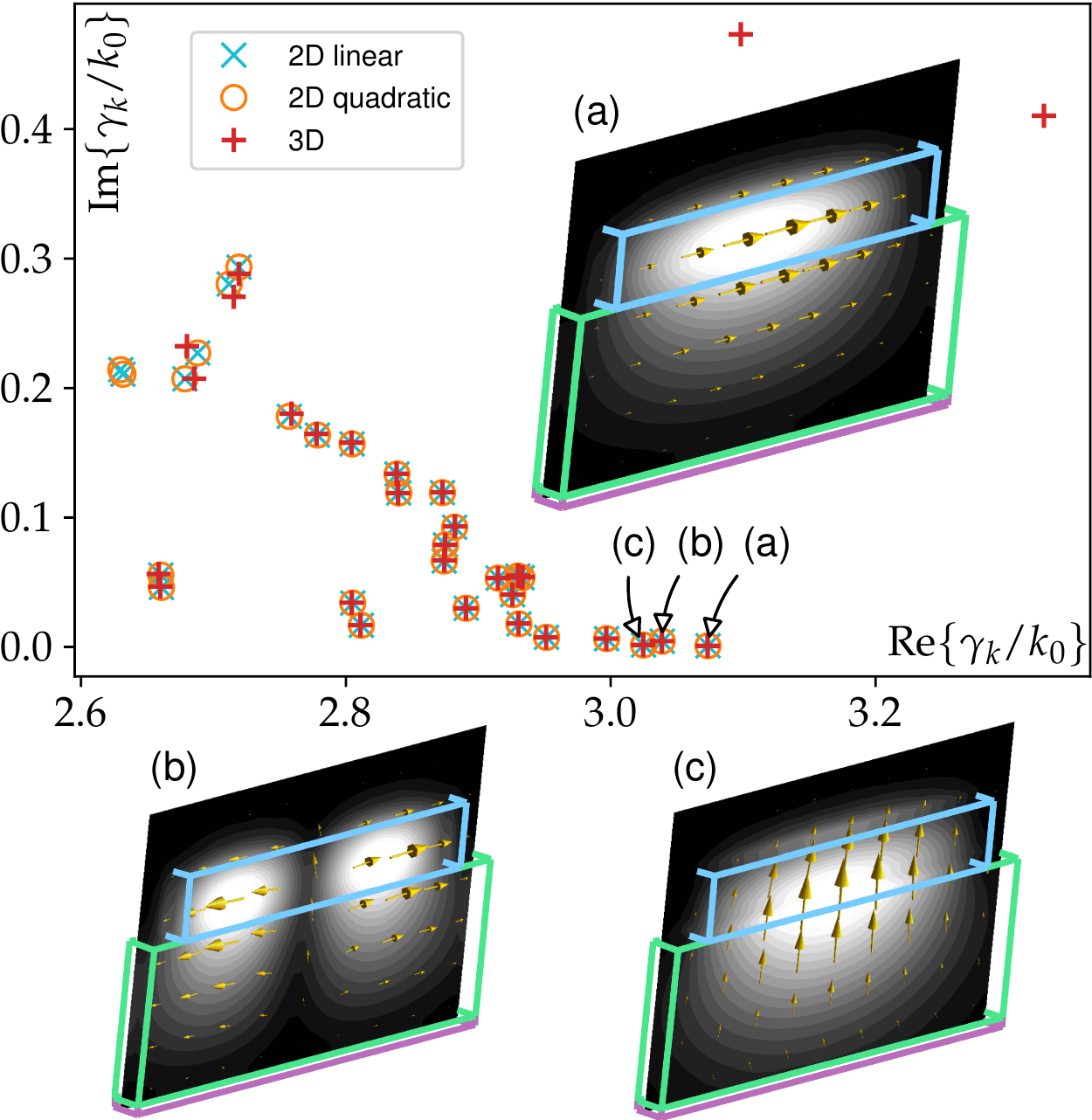} 
  \caption{{\crevii For validation of the 3D modal approach, the same structure as in Fig.~\ref{fig:modes2D} is extruded along $z$ with a distance $d=\SI{1}{\micro\meter}$. The spectrum obtained for the periodic 3D modal approach (\eqref{eq:weakmodal3D}, red pluses) is compared to the spectrum of a 2D invariant structure (blue crosses and orange circles, see the main text of Sec.~\ref{sec:3Dmodal2} for the details of the two approaches used for the genuine 2D  problem). The insets represent the three periodic parts of the modes with smallest attenuation. For each inset, two cuts are performed in the 3D domain. The first one (black and white contour plot at $z$=-$d$/4) corresponds to the norm of the  and the second one (yellow arrows at $z$=$d$/4) to the real part of the eigenvectors. Note that the modes profile in the insets correspond exactly to the modes shown in Fig.~\ref{fig:modes2D}(a,b,c).}}
  \label{fig:validation2D3D}
\end{figure}

\subsection{\crevi Discretization}
{\crevi The periodic vector unknown $\bEp$ is discretized using high order Webb hierarchical edge elements \cite{geuzaine1999convergence,webb1993hierarchal} with 26 DOFs per tetrahedron (3 DOFs per edge, 2 DOFs per face). The scalar field $\varphi$ mapping the divergence is discretized using Lagrange $P_3$ elements, with 20 DOFs per tetrahedron (4 nodal DOFs, two DOFs per edge, one DOF per face). Periodic boundary conditions are imposed along the $z$~direction for both $\bEp$ and $\varphi$.}

\subsection{Numerical validation}\label{sec:3Dmodal2}
It is apropos to validate this 3D model numerically using an extruded 2D domain. The eigenvalue resulting from three finite element problems are shown in Fig.~\ref{fig:validation2D3D}. Two of them (orange circles and blue crosses) are variants of the 2D problem in \eqref{eq:devstrongform2}. The problem is indeed quadratic and can be solved as is using the SLEPc library (orange circles) which implements its own internal numerical linearization. But as detailed in Ref.~\cite{renversez2012foundations}, it is possible to linearize the 2D problem by simply using for unknown $(e_x,e_y,i{\crevme\beta}\,e_z)$ instead of $(e_x,e_y,e_z)$. The resulting sparse systems resulting from the two methods are different and it is worth noting that the 30 eigenvalues are identical up to numerical precision. Now the third eigenproblem (red pluses in Fig.~\ref{fig:validation2D3D}) corresponds to Eqs.~(\ref{eq:weakmodal3Dcurl},\ref{eq:weakmodal3Ddiv}) applied to the 3D problem obtained by simple extrusion along $z$ of the previous 2D problem by a period $d$=\mic{1} along the $z$ direction. The value of $d$ can be arbitrarily chosen because of the translational invariance. The small value of the period along $z$ corresponds to a large first Brillouin zone so that the eigenvalues computed do not belong to a folded dispersion branch and the periodic part of the Bloch vector field is constant along $z$ as depicted in Figs.~\ref{fig:validation2D3D}(a-c). Up to the $\pi$/d folding of the dispersion curves expected from the application to the Bloch theorem, these 2D and 3D invariant problems are spectrally equivalent and the eigenvalues are indeed retrieved with excellent accuracy. The discrepancies obtained for Im$\{\gamma_k/k_0\}>0.2$ can be simply explained by the fact that the 3D mesh used in the simulation is comparatively coarser than the 2D mesh. For the modes with smallest attenuation labelled (a), (b) and (c), the relative error between the 2D and 3D eigenvalues is lower that $10^{-4}$  in modulus. Note that convergence tests have been performed and this discrepancy 
decreases with the mesh size at the expected rate given the type and order of the chosen finite elements spaces. The real parts of $\bE_{\#,k}$ are shown in the insets (a-c) of Fig.~\ref{fig:validation2D3D} (note that they cannot be directly compared to the yellow arrows of Fig.~\ref{fig:modes2D}(a-c) since an eigenvector is defined up to an arbitrary complex number).

{\crevii
Computing 30 eigenvalues of the periodic 3D structure takes \SI{80}{\second} on a laptop with low order elements (twelve DOFs per tetrahedron for the vector unknown, 10 DOFs per tetrahedron for the scalar one, about 100000~DOFs in total) with the mesh paramater $n$=4.
The same computation with $n$=8 and higher order edge elements (26~DOFs per tetrahedron for the vector unknown, twenty DOFs per tetrahedron for the scalar one, about 800000 DOFs in total) takes \SI{30}{\minute} on the 24~cores workstation.}

We are finally in position to compare the results derived from the modes of the infinitely structured waveguide (3D modal problem defined in Sec.~\ref{sec:3Dmodal}) to the transmission properties of scattering problems with a finite number of periods (3D direct problems defined in Sec.~\ref{sec:3Ddirect}).

\begin{figure}
  \centering 
  \includegraphics[width=.65\textwidth]{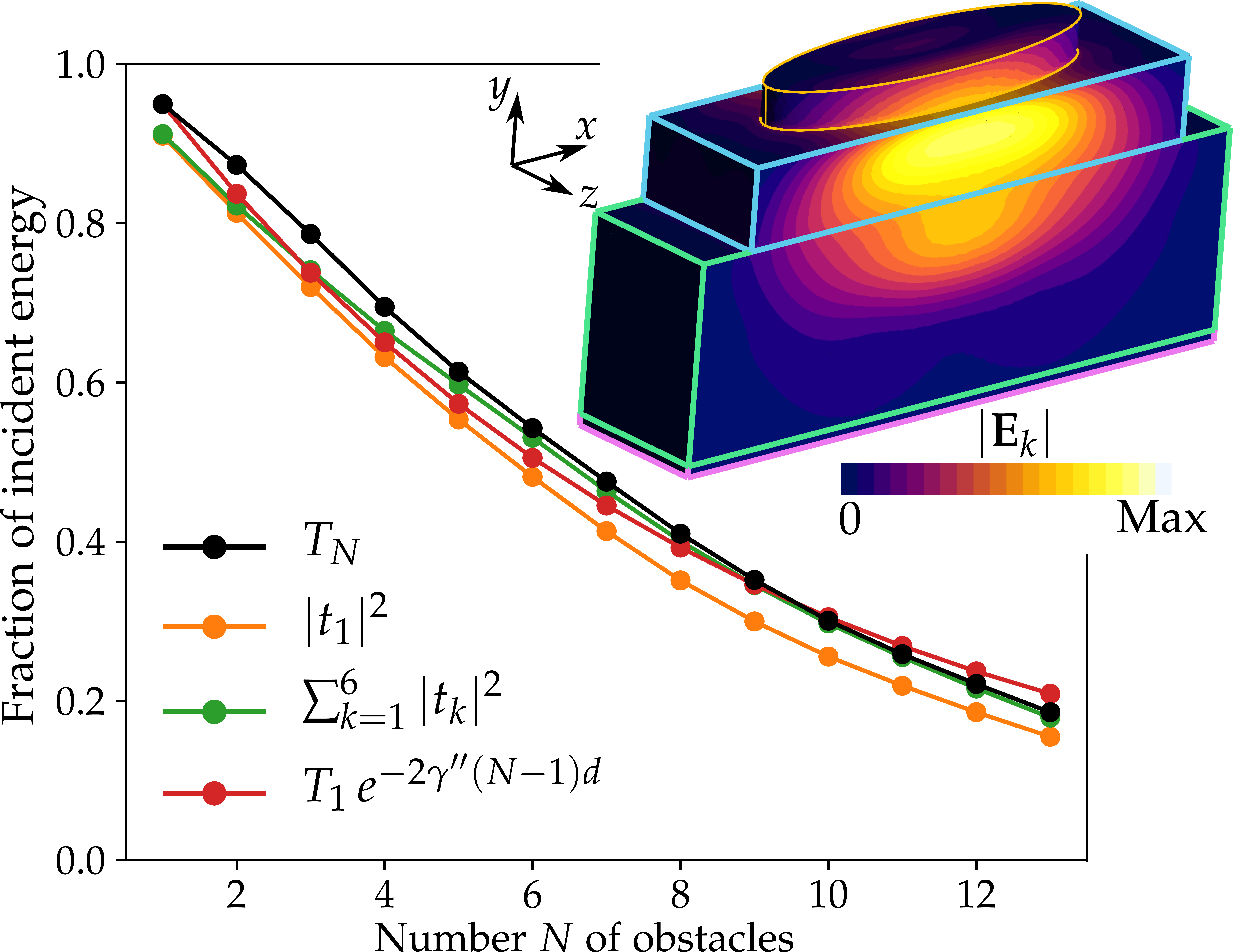}
  \caption{{\crevii The various transmission definitions (see the main text in Sec.~\ref{sec:disc}) introduced as a function of the number of obstacles. The black curve represents the transmission $T$ defined in \eqref{eq:RTA} obtained using the direct problem (each black bullet corresponds to a direct Finite Element run). The green curve corresponds to the sum of square modulus of the expansion coefficients $|t_k|$ defined in \eqref{eq:rntn}. The yellow curve shows the sole contribution of $|t_1|^2$. The red curve is obtained by solving one single 3D modal problem and shows the spatial exponential decay of the power associated with the 3D mode with lowest attenuation. The norm of the corresponding electric eigenfield is shown in at the top right corner.}}
  \label{fig:compa_modal_direct3D}
\end{figure}

\section{Discussion}
\label{sec:disc} 
The direct 3D  approach detailed in Sec.~\ref{sec:3Ddirect} and the 3D modal one based on $z$-periodicity presented in the previous section are now compared. In both cases, the unit cell of the waveguide guide contains one lossy ellipsoidal patch defined by its relative permittivity $\epsrin{d}=9+2i$, height $h=\mic{1}$, transverse and longitudinal radii $r_t=\mic{5}$ and $r_l=\mic{1.5}$, and period $d=\mic{4}$ (see~\ding{202} in Fig.~\ref{fig:scheme}). In the finite size problem, the number of patches is $N$ and the incident field is the fundamental mode labelled~1 in Fig.~\ref{fig:modes2D}(a).

Figure~\ref{fig:compa_modal_direct3D} shows the transmission $T_N$ (black curve, cf \eqref{eq:RTA}) as a function of the number $N$ unit cells (or scatterers). The orange curve represents the coefficient $|t_1|^2$ (cf \eqref{eq:rntn}), {\crevi which is a good approximation of} the fraction of incident energy carried into mode~1 and remaining in this channel after crossing the modified waveguide segment containing the scatterers. This last curve lies below the green curve that represents the sum of the amplitudes transmitted into all 6 channels (or modes) of the $z$-invariant waveguide shown in Fig.~\ref{fig:modes2D}. This numerical set up corresponds exactly to the configuration depicted in Fig.~\ref{fig:poynting} with $N=4$.

Finally, one can correlate these results to the infinitely periodic structure and superimpose the last red curve that represents {\crevme$T_1\,e^{-2\gamma''(N-1)d}$} (where $\gamma''=\mathrm{Im}\{\gamma\}$), the spatial damping of power associated with the mode with smallest propagation losses found using the modal approach detailed in Sec.~\ref{sec:3Dmodal}. Note that the normalization factor $T_1$ (transmission obtained for the direct 3D problem for one single obstacle) accounts for the fact the damping of the 3D mode computed with one obstacle does not make any sens in absence of obstacle ({\emph i.e.} for $N=0$) and represents the input impedance of the structured waveguide. {\crevii The norm of this 3D mode is represented in purple/yellow colors at the top right of the Fig.~\ref{fig:compa_modal_direct3D}. The direction of the eigenfield is not represented for clarity, but it is globally polarized along $x$. This 3D mode is very similar to mode~1 in Fig.~\ref{fig:modes2D}(a), which is the one injected in the direct problem to obtain the three other black, orange and green curves.} The consistency between these four quantities is remarkable.

\section{Conclusion} 
In conclusion, we present in this paper a general finite element frame for the study of discontinuous waveguides, from isolated discontinuities to fully periodic ones. A first method, adapted to a finite set of discontinuities allows to compute, given the modes of the invariant structure, the field scattered by the local discontinuity, all relevant energy related quantities, and the projection of the scattered field on the modes of the invariant structure (that is the elements of the transition scattering matrix). 

When the modified region extends to infinity with periodic discontinuities, the relevant quantity is the dispersion relation of the so formed structured waveguide. An adequate weak treatment of divergence condition allows to determine these modes with accuracy.

The two numerical models presented in this paper show great interest for the design of structured waveguides. Note that the methodology adopted for the  direct problem is very general and can readily be applied to a large variety of guiding structure and geometry of objects located in the modified waveguide segment.

{\crevii The two methods are in fact complementary. To give a concrete example in sensing applications, the adjunction of well chosen periodic scatterers \cite{fevrier2012giant} above a waveguide allows for instance to strengthen the interaction of the light flowing in the superstrate near the ridge where \emph{e.g.} the molecules to detect lie. In this case, it is enough to study the modes profiles of the infinitely periodic 3D structure as detailed in Sec.~\ref{sec:3Dmodal}. It is indeed much faster than optimizing the scatterers properties using the direct problem introduced in Sec.~\ref{sec:3Ddirect} applied to a very long finite chain of scatterers. However, once the properties of the scatterers optimal for the targeted application, the practical device consists indeed of a finite chain. Then, the direct problem introduced in Sec.~\ref{sec:3Ddirect} is the ideal tool to study and optimize the coupling of an incident mode into the modified segment.}

Finally, open source models allowing to retrieve most of the results of this paper are provided~\cite{code}. They can be tuned to handle different geometries and material properties.

This work will later be extended to the case where the input and output invariant structures mismatch using a coupled mode-FE approach~\cite{pelat2011coupled} to compute the relevant incident fields.

\section*{Acknowledgement}
This research was supported by ANR Louise project, grant ANR-15-CE04-000164 of the French Agence Nationale de la Recherche. The authors would like to thank the developers of MUMPS~\cite{mumps-userguide}, PETSc~\cite{petsc-user-ref}, SLEPc~\cite{Hernandez2005-SSF}, Gmsh~\cite{gmsh} and GetDP~\cite{getdp} for maintaining and making their respective libraries freely available. Finally, the authors acknowledge Sonia Fliss (INRIA POEMS) for fruitful discussions.
\section*{Disclosures}
The authors declare no conflicts of interest.

\bigskip
\bibliographystyle{ieeetr}
\bibliography{biblio-guide_louise}

\begin{thebibliography}{10}

\bibitem{schinger68discontinuities-waveguides}
J.~Schwinger and D.~S. Saxon, {\em Discontinuities in waveguides}.
\newblock Gordon and Breach, 1968.

\bibitem{lewin75theory-waveguides}
L.~Lewin, {\em Theory of waveguides}.
\newblock Newnes-Butter-worths, 1975.

\bibitem{taflove05}
A.~Taflove and S.~C. Hagness, {\em Computational Electrodynamics: The
  Finite-Difference Time-Domain Method}.
\newblock Artech House, Boston, 3~ed., 2005.

\bibitem{taflove13}
A.~Taflove, A.~Oskooi, and S.~G. Johnson, eds., {\em Advances in {FDTD}
  Computational Electrodynamics}.
\newblock Photonics and Nanotechnology, Artech House, Boston, 2013.

\bibitem{lalanne2000fourier}
P.~Lalanne and E.~Silberstein, ``Fourier-modal methods applied to waveguide
  computational problems,'' {\em Optics Letters}, vol.~25, no.~15,
  pp.~1092--1094, 2000.

\bibitem{lecamp2007theoretical}
G.~Lecamp, J.-P. Hugonin, and P.~Lalanne, ``Theoretical and computational
  concepts for periodic optical waveguides,'' {\em Optics express}, vol.~15,
  no.~18, pp.~11042--11060, 2007.

\bibitem{pagneux1996study}
V.~Pagneux, N.~Amir, and J.~Kergomard, ``A study of wave propagation in varying
  cross-section waveguides by modal decomposition. {P}art {I}. {T}heory and
  validation,'' {\em The Journal of the Acoustical Society of America},
  vol.~100, no.~4, pp.~2034--2048, 1996.

\bibitem{pelat2011coupled}
A.~Pelat, S.~Felix, and V.~Pagneux, ``A coupled modal-finite element method for
  the wave propagation modeling in irregular open waveguides,'' {\em The
  Journal of the Acoustical Society of America}, vol.~129, no.~3,
  pp.~1240--1249, 2011.

\bibitem{rosenkrantz2018benchmarking}
J.~R. de~Lasson, L.~H. Frandsen, P.~Gutsche, S.~Burger, O.~S. Kim,
  O.~Breinbjerg, A.~Ivinskaya, F.~Wang, O.~Sigmund, T.~H\"{a}yrynen, A.~V.
  Lavrinenko, J.~M{\o}rk, and N.~Gregersen, ``{\crevii Benchmarking five
  numerical simulation techniques for computing resonance wavelengths and
  quality factors in photonic crystal membrane line defect cavities},'' {\em
  Opt. Express}, vol.~26, pp.~11366--11392, Apr 2018.

\bibitem{lalanne2019quasinormal}
P.~Lalanne, W.~Yan, A.~Gras, C.~Sauvan, J.-P. Hugonin, M.~Besbes,
  G.~Dem\'{e}sy, M.~D. Truong, B.~Gralak, F.~Zolla, A.~Nicolet, F.~Binkowski,
  L.~Zschiedrich, S.~Burger, J.~Zimmerling, R.~Remis, P.~Urbach, H.~T. Liu, and
  T.~Weiss, ``{\crevii Quasinormal mode solvers for resonators with dispersive
  materials},'' {\em J. Opt. Soc. Am. A}, vol.~36, pp.~686--704, Apr 2019.

\bibitem{jin02FEM-electromag}
J.~Jin, {\em The {F}inite {E}lement {M}ethod in {E}lectromagnetics}.
\newblock John Wiley \& Sons Inc., 3rd~ed., 2014.

\bibitem{bendoe-sigmund04topo-opti}
M.~P. Bendsoe and O.~Sigmund, {\em Topology Optimization}.
\newblock Springer-Verlag, 2nd~ed., 2004.

\bibitem{gutierrez2016optical}
A.~Gutierrez-Arroyo, E.~Baudet, L.~Bodiou, J.~Lemaitre, I.~Hardy, F.~Faijan,
  B.~Bureau, V.~Nazabal, and J.~Charrier, ``Optical characterization at 7.7
  {$\mu$}m of an integrated platform based on chalcogenide waveguides for
  sensing applications in the mid-infrared,'' {\em Opt. Express}, vol.~24,
  p.~23109, 2016.

\bibitem{Desevedavy:09}
F.~D\'{e}s\'{e}v\'{e}davy, G.~Renversez, J.~Troles, L.~Brilland, P.~Houizot,
  Q.~Coulombier, F.~Smektala, N.~Traynor, and J.-L. Adam, ``{Te-As-Se} glass
  microstructured optical fiber for the middle infrared,'' {\em Appl. Opt.},
  vol.~48, pp.~3860--3865, Jul 2009.

\bibitem{eggleton2011chalcogenide}
B.~J. Eggleton, B.~Luther-Davies, and K.~Richardson, ``Chalcogenide
  photonics,'' {\em Nature photonics}, vol.~5, no.~3, p.~141, 2011.

\bibitem{nedelec91}
T.~Nédélec, {\em Notions sur les techniques d'éléments finis,
  mathématiques et applications, n$^{o}$ 7}.
\newblock Mathématiques \& Applications, Ellipses, 2nd~ed., 1991.

\bibitem{renversez2012foundations}
F.~Zolla, G.~Renversez, A.~Nicolet, B.~Kuhlmey, S.~Guenneau, D.~Felbacq,
  A.~Argyros, and S.~Leon-Saval, {\em Foundations of Photonic Crystal Fibres}.
\newblock Imperial College Press, 2nd~ed., 2012.

\bibitem{demesy2010all}
G.~Dem{\'e}sy, F.~Zolla, A.~Nicolet, and M.~Commandr{\'e}, ``All-purpose finite
  element formulation for arbitrarily shaped crossed-gratings embedded in a
  multilayered stack,'' {\em JOSA A}, vol.~27, no.~4, pp.~878--889, 2010.

\bibitem{snyder1983optical}
A.~Snyder and J.~Love, {\em Optical Waveguide Theory Chapman and Hall}, p.~500.
\newblock New York, first~ed., 1983.

\bibitem{nicolet2004modelling}
A.~Nicolet, S.~Guenneau, C.~Geuzaine, and F.~Zolla, ``Modelling of
  electromagnetic waves in periodic media with finite elements,'' {\em Journal
  of Computational and Applied Mathematics}, vol.~168, no.~1-2, pp.~321--329,
  2004.

\bibitem{kuriakose17Measurement-ultrafast-optical-Kerr-effect-chalco}
T.~Kuriakose, E.~Baudet, T.~Halenkovič, M.~M. Elsawy, P.~Němec, V.~Nazabal,
  G.~Renversez, and M.~Chauvet, ``Measurement of ultrafast optical kerr effect
  of ge-sb-se chalcogenide slab waveguides by the beam self-trapping
  technique,'' {\em Opt. Comm.}, vol.~403, pp.~352 -- 357, 2017.

\bibitem{chandler2005high}
D.~Chandler-Horowitz and P.~M. Amirtharaj, ``High-accuracy, midinfrared
  (450~cm$^{-1}\leqslant\omega\leqslant 4000\,$cm$^{-1}$) refractive index
  values of silicon,'' {\em Journal of Applied physics}, vol.~97, no.~12,
  p.~123526, 2005.

\bibitem{gmsh}
C.~Geuzaine and J.~F. Remacle, ``Gmsh: a three-dimensional finite element mesh
  generator with built-in pre- and post-processing facilities,'' {\em
  International Journal for Numerical Methods in Engineering}, vol.~79, no.~11,
  pp.~1309--1331, 2009.

\bibitem{getdp}
{P. Dular, C. Geuzaine, F. Henrotte and W. Legros}, ``A general environment for
  the treatment of discrete problems and its application to the finite element
  method,'' {\em IEEE Transactions on Magnetics}, vol.~34, no.~5,
  pp.~3395--3398, 1998.

\bibitem{code}
G.~Demésy and G.~Renversez,
  ``\url{https://gitlab.fresnel.fr/fem_models/structured_waveguides},'' 2019.

\bibitem{zolla1996method}
F.~Zolla and R.~Petit, ``Method of fictitious sources as applied to the
  electromagnetic diffraction of a plane wave by a grating in conical
  diffraction mounts,'' {\em JOSA A}, vol.~13, no.~4, pp.~796--802, 1996.

\bibitem{berenger94perfec-match-layer}
J.-P. Bérenger, ``A perfectly matched layer for the absorption of
  electromagnetic waves,'' {\em Journal of Computational Physics}, vol.~114,
  pp.~185--200, 1994.

\bibitem{Hernandez2005-SSF}
V.~Hernandez, J.~E. Roman, and V.~Vidal, ``{SLEPc}: A scalable and flexible
  toolkit for the solution of eigenvalue problems,'' {\em {ACM} Trans. Math.
  Software}, vol.~31, no.~3, pp.~351--362, 2005.

\bibitem{geuzaine1999convergence}
C.~Geuzaine, B.~Meys, P.~Dular, and W.~Legros, ``{\crevii Convergence of high
  order curl-conforming finite elements [for {EM} field calculations]},'' {\em
  IEEE Transactions on Magnetics}, vol.~35, no.~3, pp.~1442--1445, 1999.

\bibitem{webb1993hierarchal}
J.~Webb and B.~Forgahani, ``{\crevii Hierarchal scalar and vector
  tetrahedra},'' {\em IEEE Transactions on Magnetics}, vol.~29, no.~2,
  pp.~1495--1498, 1993.

\bibitem{mumps-userguide}
P.~Amestoy, I.~Duff, A.~Guermouche, J.~Koster, J.-Y. L'Excellent, and
  S.~Pralet, {\em MUltifrontal Massively Parallel Solver, (MUMPS 4.8.4), Users'
  guide}.
\newblock CERFACS, ENSEEIHT-IRIT, and INRIA, December 2008.
\newblock {h}ttp://mumps.enseeiht.fr and http://graal.ens-lyon.fr/MUMPS.

\bibitem{sammut1976leaky}
R.~Sammut and A.~W. Snyder, ``{\crevi Leaky modes on a dielectric waveguide:
  orthogonality and excitation},'' {\em Appl. Opt.}, vol.~15, pp.~1040--1044,
  Apr 1976.

\bibitem{joannopoulos95photon-crist-modlin-flow-light}
J.~D. Joannopoulos, R.~Meade, and J.~N. Winn, {\em Photonic Crystals Molding
  the Flow of Light}.
\newblock Princeton University Press, 1995.

\bibitem{lackner2019}
C.~Lackner, S.~Meng, and P.~Monk, ``Determination of electromagnetic bloch
  variety in a medium with frequency-dependent coefficients,'' {\em Journal of
  Computational and Applied Mathematics}, vol.~358, pp.~359 -- 373, 2019.

\bibitem{fevrier2012giant}
M.~F{\'e}vrier, P.~Gogol, A.~Aassime, R.~M{\'e}gy, C.~Delacour, A.~Chelnokov,
  A.~Apuzzo, S.~Blaize, J.-M. Lourtioz, and B.~Dagens, ``Giant coupling effect
  between metal nanoparticle chain and optical waveguide,'' {\em Nano letters},
  vol.~12, no.~2, pp.~1032--1037, 2012.

\bibitem{petsc-user-ref}
S.~Balay, K.~Buschelman, V.~Eijkhout, W.~D. Gropp, D.~Kaushik, M.~G. Knepley,
  L.~C. McInnes, B.~F. Smith, and H.~Zhang, {\em {PETS}c Users Manual}, 2007.

\end{thebibliography}





\end{document}